\documentclass[preprint,aps,superscriptaddress,showpacs,floatfix]{revtex4}
\usepackage{graphicx}
\begin{document}
\title{Pentaquark baryon production from photon-neuteron reactions}
\bigskip
\author{W. Liu}
\affiliation{Cyclotron Institute and Physics Department, Texas A$\&$M
University, College Station, Texas 77843-3366, USA}
\author{C. M. Ko}
\affiliation{Cyclotron Institute and Physics Department, Texas A$\&$M
University, College Station, Texas 77843-3366, USA}

\date{\today}

\begin{abstract}
Extending the hadronic Lagrangians that we recently introduced for
studying pentaquark $\Theta^+$ baryon production from meson-proton,
proton-proton, and photon-proton reactions near threshold to include
the anomalous interaction between $\gamma$ and $K^*K$, we evaluate
the cross section for $\Theta^+$ production from photon-neutron
reactions, in which the $\Theta^+$ was first detected in the SPring-8 
experiment in Japan and the CLAS experiment at Thomas Jefferson
National Laboratory. With empirical coupling constants and form factors, 
and assuming that the decay width of $\Theta^+$ is 20 MeV, the predicted 
cross section is found to have a peak value of about 280 nb, which is 
substantially larger than that for $\Theta^+$ production from 
photon-proton reactions.

\end{abstract}


\maketitle

\section{introduction}
Recently, a narrow baryon state was inferred from the invariant mass
spectrum of $K^+n$ or $K^0p$ in nuclear reactions induced by photons
\cite{nakano,stepanyan,barth} or kaons \cite{barmin}. The extracted mass of
about 1.54 GeV and width of less than 21-25 MeV are consistent with those
of the pentaquark baryon $\Theta^+$ consisting of $uudd\bar s$ quarks
and predicted in the chiral soliton model \cite{diakonov}. Its existence
has also been verified recently in the constituent quark model
\cite{riska,lipkin} and the QCD sum rules \cite{zhu}. Although the
spin and isospin of $\Theta^+$ are predicted to be 1/2 and 0,
respectively, those of the one detected in experiments are not yet determined.
Studies have therefore been carried out to predict its decay branching
ratios based on different assignments of its spin and isospin \cite{carl,ma}.
To evaluate the cross sections for $\Theta^+$ production from nucleons
induced by mesons and protons, which are needed for studying $\Theta+$
production in relativistic heavy ion collisions \cite{chen},
we have used a hadronic model that is based on $SU(3)$ flavor symmetry 
with empirical hadron masses and form factors at interaction vertices 
\cite{li,lin,di}. For the coupling constant between $\Theta^+$ and 
$KN$, it is determined from the width of $\Theta^+$. With the photon
included as a $U_{\rm em}(1)$ gauge boson, we have further generalized 
the hadronic model to calculate the cross sections for $\Theta^+$ 
production from photon-proton reactions. We find that the predicted 
cross sections have peak values about 1.5 mb in kaon-proton reactions, 
0.1 mb in rho-nucleon reactions, 0.05 mb in pion-nucleon reactions, 
20 $\mu$b in proton-proton reactions, and 40 nb in photon-proton 
reactions \cite{liuko}.

In the present paper, we extend this model to include the anomalous
interaction between photon and $K^*K$ and to use it to evaluate
the cross section for $\Theta^+$ production from photon-neutron
reactions, in which the $\Theta^+$ baryon was first detected in
the SPring-8 experiment in Japan and the CLAS experiment at Thomas
Jefferson National Laboratory.

This paper is organized as follows. In Sect. \ref{photonneutron}, the cross
section for $\Theta^+$ production from photon-neutron reactions with 
inclusion of final states up to four particles is evaluated. The process 
involving the coupling of $\Theta^+$ to $K^*N$ in the reaction 
$\gamma n\to K^-\Theta^+$ is discussed in Sect. \ref{discussions}. 
Finally, a summary is given in Sect. \ref{summary}.

\section{Theta production from photon-neutron reactions}\label{photonneutron}

In photon-neutron interactions, the $\Theta^+$ baryon can be produced in
various final states. In the present study, we consider reactions with
final states up to four particles.  For final states with two particles, the
reaction is $\gamma n\to K^-\Theta^+$ and is described in
Sect.\ref{twobody}. In Sect.\ref{threebody}, the reactions
$\gamma n\to\pi^0 K^-\Theta^+$ and $\gamma n\to\pi^-\bar K^0\Theta^+$
with three particles in the final state are discussed. The reactions
$\gamma n\to\pi^-\bar K^{*0}\Theta^+$,
$\gamma n\to\pi^0K^{*-}\Theta^+$, $\gamma n\to\rho^-\bar K^{*0}\Theta^+$,
and $\gamma n\to\rho^0K^-\Theta^+$ with four particles in the final state,
if we take into account decays of $K^*$ to $K\pi$ and $\rho$ to $\pi\pi$,
are given in Sect. \ref{fourbody}.  Numerical results for the cross
sections for these reactions are presented in Sect.\ref{results}.

\subsection{$\gamma n\to K^-\Theta^+$}\label{twobody}

\begin{figure}[ht]
\includegraphics[width=3.0in,height=1.5in,angle=0]{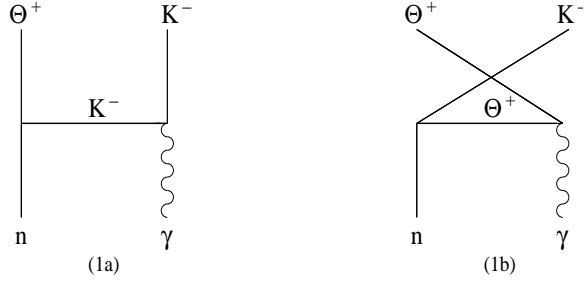}
\caption{Diagrams for $\Theta^+$ production from photon-neutron
reactions with two-body final states.} \label{diagram1}
\end{figure}

For the reaction $\gamma n\to K^-\Theta^+$ with two-body final state,
the two diagrams are shown in Fig.\ref{diagram1}. To evaluate its cross
section, we need the following interaction Lagrangians:
\begin{eqnarray}
{\cal L}_{KN\Theta}&=&ig_{KN\Theta}(\bar\Theta\gamma_5N\bar K
+\bar N\gamma_5 \Theta K),\nonumber\\
{\cal L}_{\gamma \Theta\Theta}&=&ieA^\mu\bar \Theta\gamma_\mu\Theta,\nonumber\\
{\cal L}_{\gamma KK}&=&ieA^\mu[KQ\partial_\mu\bar K-\partial_\mu KQ\bar K],
\end{eqnarray}
In the above, $\Theta$ denotes the spin 1/2 and isospin 0 pentaquark baryon;
$N$ and $K$ are, respectively, the isospin doublet kaon and nucleon
fields; and $A_\mu$ is the photon field. The operator $Q$ is the
diagonal charge operator with elements 0 and -1. The coupling constant
$g_{KN\Theta}$ of $\Theta^+$ to $NK$ is determined from its decay width
given by
\begin{equation}
\Gamma_\Theta=\frac{g_{KN\Theta}^2}{2\pi}\frac{k(\sqrt{m_N^2+k^2}-m_N)}
{m_\Theta},
\end{equation}
where $m_N$ and $m_\Theta$ are masses of nucleon and $\Theta^+$,
respectively, and $k$ is the momentum of $N$ or $K$ in the rest frame
of $\Theta^+$. Using $m_\Theta=1.54$ GeV and $\Gamma_\Theta=20$ MeV,
we find $g_{KN\Theta}=4.4$, which is comparable to that given by the
chiral soliton model \cite{diakonov}.

The two amplitudes for the reaction $\gamma n\to K^-\Theta^+$ are then
given by
\begin{eqnarray}
{\cal M}_{1a}&=&-eg_{KN\Theta}\frac{F({\bf q}^2)}{t-m^2_K}
(p_2-2p_4)^\mu\bar\Theta(p_3)\gamma_5 n(p_1)\epsilon_\mu,\nonumber\\
{\cal M}_{1b} & = &-eg_{KN\Theta}\frac{F({\bf q}^2)}{u-m^{2}_{\Theta}}
\bar{\Theta}(p_3)\gamma^\mu({p\mkern-8mu/}_1-{p\mkern-8mu/}_4
+m_{\Theta})\gamma_5 n(p_1)\epsilon_\mu,
\end{eqnarray}
with $\epsilon_\mu$ denoting the photon polarization vector. In the above,
$p_1$ and $p_2$ are momenta of $n$ and $\gamma$ while $p_3$ and $p_4$ are 
those of $\Theta^+$ and $K^-$, respectively. The usual Mandelstam
variables are given by $s=(p_1+p_2)^2$, $t=(p_1-p_3)^2$, and 
$u=(p_1-p_4)^2$ with $s$ needed later. To take into account the final 
sizes of hadrons and also maintain the gauge invariance of the total
amplitudes, we have included the same overall form factor $F({\bf q}^2)$  
in each amplitude as in our previous study of $\Theta^+$ production
from photon-proton reactions. It is taken to have the following
monopole form
\begin{eqnarray}
F({\bf q}^2)=\frac{\Lambda^2}{\Lambda^2+{\bf q}^2}
\label{form},
\end{eqnarray}
with ${\bf q}$ being the photon three momentum in center-of-mass system.
The cutoff parameter $\Lambda$ is taken to be $\Lambda=0.75$ GeV, which
has been shown to reproduce the experimental data at center-of-mass
energy of 6 GeV using similar Lagrangians based on the SU(4) flavor
symmetry with empirical masses and coupling constants \cite{liu2}.

The cross section for the reaction $\gamma p\to K^-\Theta^+$ with
two-body final state is then given by 
\begin{eqnarray}\label{cross}
\frac{d\sigma}{dt}=\frac{1}{256\pi s{\bf q}^2}
\sum_{spin}|{\cal M}_{1a}+{\cal M}_{1b}|^2,
\end{eqnarray}
with the summation over both initial neutron and final $\Theta^+$ spins.

\subsection{$\gamma n\to\pi^0 K^-\Theta^+$ and
$\gamma n\to\pi^-\bar K^0\Theta^+$}\label{threebody}

\begin{figure}[ht]
\includegraphics[width=3.0in,height=3.0in,angle=0]{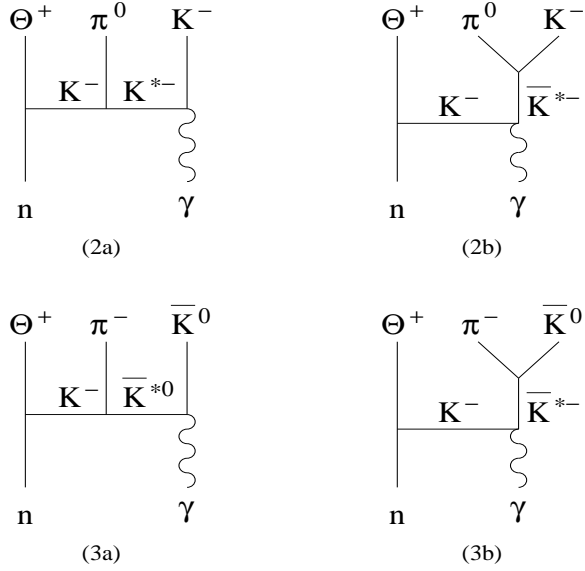}
\caption{Diagrams for $\Theta^+$ production from photon-proton
reactions with three-body final states.} \label{diagram2}
\end{figure}

For the reactions $\gamma n\to\pi^0 K^-\Theta^+$ and
$\gamma n\to\pi^-\bar K^0\Theta^+$ with three-body final states, their
diagrams are shown in Fig.\ref{diagram2}. These reactions are due to the
the anomalous parity interaction between photon and $KK^*$, i.e.,
\begin{eqnarray}
{\cal L}_{\gamma KK^{*0}}=g_{\gamma KK^*}\epsilon_{\alpha\beta\mu\nu}
\partial^\alpha A^\beta[\partial^\mu\bar K^{*\nu}K+\bar K\partial^\mu
K^{*\nu}],
\end{eqnarray}
where $\epsilon_{\alpha\beta\mu\nu}$ is the antisymmetric tensor
with $\epsilon_{0123}=1$. The coupling constant $g_{\gamma KK^*}$
has the dimension of inverse of energy and can be determined from the
decay width of $K^{*0}$ or $K^{*\pm}$ to photons through the process
$K^{*0}\to K^0\gamma$ or $K^{*\pm}\to K^\pm\gamma$, respectively, i.e.,
\begin{eqnarray}
\Gamma_{K^*}=\frac{g^2_{\gamma KK^*}}{12\pi}k^3,
\end{eqnarray}
with $k$ being photon momentum in the rest frame of $K^*$.
From the empirical values 
$\Gamma_{K^{*0}\to K^0\gamma}=50.8\times 2.3\times 10^{-3}=0.117$ MeV and 
$\Gamma_{K^{*\pm}\to K^\pm\gamma}=50.8\times 9.9\times 10^{-4}=0.05$ MeV, 
where the first factor 50.8 MeV is the $K^*$ total decay width while
the second factor is the branching ratio for electromagnetic decay, 
we find that $g_{\gamma K^0K^{*0}}=0.0388$ GeV$^{-1}$ and 
$g_{\gamma K^\pm K^{*\pm}}=0.0257$ GeV$^{-1}$.

Also needed is the interaction Lagrangian involving $\pi$, $K$, and $K^*$,
which is given by
\begin{equation}
{\cal L}_{\pi KK^*}=ig_{\pi KK^*}K^{*\mu}(\bar K\partial_\mu\pi
-\partial_\mu\bar K\pi)+{\rm H.c.},
\end{equation}
with the coupling constant $g_{\pi KK^*}=3.28$ determined from the
decay width of $K^*$ to $K$ and $\pi$ \cite{ko}.

The amplitudes for the reactions $\gamma n\to\pi^0 K^-\Theta^+$
and $\gamma n\to\pi^-\bar K^0\Theta^+$ are then given by
\begin{eqnarray}
{\cal M}_i&=&ig_{KN\Theta}\bar\Theta(p_3)\gamma_5n(p_1)
\frac{F({\bf q}^2)}{t-m^2_K}
[F(({\bf p_1-p_3})^2){\cal M}^{sub}_{ia}+{\cal M}^{sub}_{ib}],
\end{eqnarray}
where
\begin{eqnarray}
{\cal M}^{sub}_{2a}&=& \frac{g_{\gamma K^-K^{*-}}g_{\pi KK^*}} 
{(k_1-k_3)^2-m^2_{K^*}+im_{K^*}\Gamma_{K^*}}
\epsilon_{\alpha\beta\mu\nu}p^\alpha_2
\epsilon^\beta_2(k_3-k_1)^\mu(k_1+k_3)^\nu,\nonumber\\
{\cal M}^{sub}_{2b}&=&\frac{g_{\gamma K^-K^{*-}}g_{\pi KK^*}} 
{(k_1+k_2)^2-m^2_{K^*}+im_{K^*}\Gamma_{K^*}}
\epsilon_{\alpha\beta\mu\nu}p^\alpha_2
\epsilon^\beta_2(k_1+k_2)^\mu(k_3-k_4)^\nu, \nonumber\\
{\cal M}^{sub}_{3a}&=& \frac{\sqrt{2}g_{\gamma K^0K^{*0}}g_{\pi KK^*}}
{(k_1-k_3)^2-m^2_{K^*}+im_{K^*}\Gamma_{K^*}}
\epsilon_{\alpha\beta\mu\nu}p^\alpha_2
\epsilon^\beta_2(k_3-k_1)^\mu(k_1+k_3)^\nu,\nonumber\\
{\cal M}^{sub}_{3b}&=&\frac{\sqrt{2}g_{\gamma K^-K^{*-}}g_{\pi KK^*}}
{(k_1+k_2)^2-m^2_{K^*}+im_{K^*}\Gamma_{K^*}}
\epsilon_{\alpha\beta\mu\nu}p^\alpha_2
\epsilon^\beta_2(k_1+k_2)^\mu(k_3-k_4)^\nu,
\end{eqnarray}
are amplitudes for the subprocesses $\gamma K^-\to\pi^0K^-$
and $\gamma K^-\to\pi^-\bar K^0$. In the above, we
have included the width of $K^*$ in its propagator to avoid possible
singularity when it becomes on shell. Also, besides the overall
monopole form factor $F({\bf q}^2)$, an additional monopole form factor
$F(({\bf p_1-p_3})^2)$ depending on the three momentum of $K^-$
and with cutoff parameter $\Lambda=0.5$ GeV, which is based on 
fitting the measured cross section for the reaction $\pi N\to K\Lambda$ 
using similar Lagrangians \cite{liu3}, has been included at the 
$KN\Theta$ vertex in diagrams (2a) and (3a). Since diagrams (2b) and 
(3b) of Fig.\ref{diagram2} can also be viewed as the reaction 
$\gamma n\to K^{*-}\Theta^+$ with two-body final state, 
their cross sections thus involve only the overall form factor 
$F({\bf q}^2)$ as in reactions with two-body final state.

The cross sections for reactions with three-body final state can
be expressed in terms of the off-shell cross sections for their 
two-body subprocesses
\cite{liu1,liu3}. For the reaction $\gamma n\to\pi^0K^-\Theta^+$,
its cross section is thus 
\begin{eqnarray}\label{32cross}
\frac{d\sigma_{\gamma n\to\pi^0 K^-\Theta^+}}{dtds_1}=
\frac{g^{2}_{KN\Theta}}{32\pi^{2}s{\bf q}^2}k\sqrt{s_{1}}
[-t+(m_N-m_{\Theta})^2]\frac{F^2({\bf q}^2)}{(t-m^{2}_{K})^{2}}
\sigma_{\gamma K^-\to\pi^0K^-}(s_{1},t),
\end{eqnarray}
and similarly for the reaction $\gamma p\to\pi^0\bar K^0\Theta^+$.
In the above, $s_1$ and $k$ are the squared center-of-mass energy
and three momentum of virtual $K^-$ in the subprocess
$\gamma K^-\to\pi^0K^-$. The cross section 
$\sigma_{\gamma K^-\to\pi^0K^-}(s_{1},t)$ for this off-shell
subprocess is calculated with the amplitude 
$F(({\bf p_1-p_3})^2){\cal M}^{sub}_{2a}+{\cal M}^{sub}_{2b}$.

\subsection{$\gamma n\to\pi^-\bar K^{*0}\Theta^+$,
$\gamma n\to\pi^0K^{*-}\Theta^+$, $\gamma n\to\rho^-\bar K^{*0}\Theta^+$,
and $\gamma n\to\rho^0K^-\Theta^+$}\label{fourbody}

\begin{figure}[ht]
\includegraphics[width=4.5in,height=6in,angle=0]{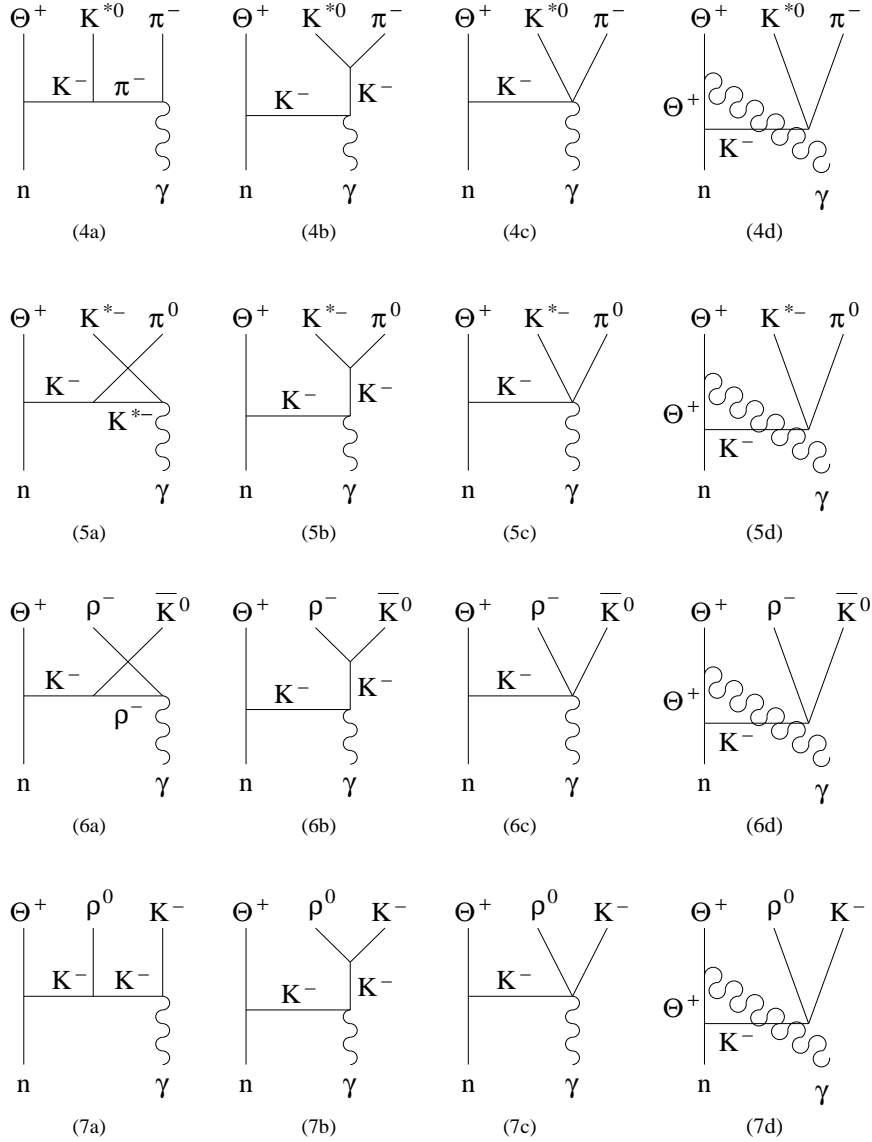}
\caption{Diagrams for $\Theta^+$ production from photon-neutron
reactions with four-body final states after taking into account the
decay of rho meson to two pions and of $K^*$ to $K$ and $\pi$.}
\label{diagram3}
\end{figure}

The reactions $\gamma n\to\pi^-\bar K^{*0}\Theta^+$,
$\gamma n\to\pi^0K^{*-}\Theta^+$, $\gamma n\to\rho^-\bar K^{*0}\Theta^+$,
and $\gamma n\to\rho^0K^-\Theta^+$ shown in Fig.\ref{diagram3} involve
four-body final states after taking into account the decay of rho meson
to two pions and of $K^*$ to $K$ and $\pi$.  In these reactions,
the photon couples either to $K$ meson as in first three diagrams or to
$\Theta^+$ as in the last diagram. Since the contribution from the
latter diagram is expected to be much smaller than those from the
former diagrams due to additional baryon propagator, as shown explicitly
in charmed hadron production from photon-proton reactions with three-body
final states \cite{liu2}, they are neglected in following calculations.
As a result, results obtained in present study for $\Theta^+$ production
with four-body final states violate slightly the gauge invariance.

To evaluate the diagrams in Fig.\ref{diagram3}, we also need the
following interaction Lagrangians:
\begin{eqnarray}
{\cal L}_{\gamma\pi\pi}&=&eA^\mu(\partial_\mu\vec\pi\times\vec\pi)_3,
\nonumber\\
{\cal L}_{\gamma \rho\rho} & = & e\{A^{\mu}
(\partial_{\mu}\vec\rho^{\nu}\times\vec{\rho}_{\nu})_3
+[(\partial_{\mu}A^\nu \vec\rho_{\nu}-A^{\nu}
\partial_{\mu}\vec\rho_{\nu})\times\vec\rho^\mu]_3+[\vec\rho^\mu\times(A^{\nu}
\partial_{\mu}\vec\rho_\nu-\partial_{\mu}A^{\nu}\vec\rho_{\nu})]_3\},
\nonumber\\
{\cal L}_{\gamma\pi KK^{*}} & = & -e g_{\pi KK^{*}}A^{\mu}(K^{*}_{\mu}
(2\vec{\tau}Q-Q\vec{\tau}) \bar{K}+K(2Q\vec{\tau}-\vec{\tau} Q)
\bar{K}^{*}_{\mu})\cdot\vec\pi,\nonumber\\
{\cal L}_{\gamma\rho KK} & = & e g_{\rho KK}A^{\mu}K(\vec{\tau}Q+Q 
\vec{\tau})\bar{K}\cdot\vec{\rho}_{\mu},
\end{eqnarray}
where the isospin triplet pion and rho meson fields are given by
$\pi=\vec\tau\cdot\vec\pi$ and $\rho^\mu=\vec\tau\cdot\vec\rho^\mu$,
respectively, with $\vec\tau$ denoting the Pauli spin matrices.
The coupling constant $g_{\rho KK}=3.25$ is obtained from the
empirical $\rho NN$ coupling via $SU(3)$ relations \cite{li}.

As in reactions with three-body final state shown in Fig.\ref{diagram2},
the amplitudes for the four reactions shown in Fig.\ref{diagram3} can
be written as
\begin{eqnarray}
{\cal M}_i&=&ig_{KN\Theta}\bar\Theta(p_3)\gamma_5n(p_1)
\frac{F({\bf q}^2)F(({\bf p_1-p_3})^2)}{t-m^2_K}
({\cal M}^{sub}_{ia}+{\cal M}^{sub}_{ib}+{\cal M}^{sub}_{ic}).
\label{four1}
\end{eqnarray}
The amplitudes ${\cal M}^{sub}_{ia}$, ${\cal M}^{sub}_{ib}$,
and ${\cal M}^{sub}_{ic}$ are for the subprocesses,
and they are given explicitly by
\begin{eqnarray}
{\cal M}^{sub}_{4a}&=&\sqrt{2}eg_{\pi KK^*}(-2k_1+k_3)^\mu
\frac{1}{(k_1-k_3)^2-m^2_\pi}(k_1 -k_3 +k_4)^\nu\varepsilon_{3\mu}
\varepsilon_{2\nu},\nonumber\\
{\cal M}^{sub}_{4b}&=&-\sqrt{2}eg_{\pi KK^*}(2k_1+k_2)^\nu
\frac{1}{(k_1+k_2)^2-m^2_K}(k_1+k_2+k_4)^\mu\varepsilon_{3\mu}
\varepsilon_{2\nu},\nonumber\\
{\cal M}^{sub}_{4c}&=&2\sqrt{2}eg_{\pi KK^*}g^{\mu\nu}\varepsilon_{3\mu}
\varepsilon_{2\nu},\nonumber\\
{\cal M}^{sub}_{5a}&=&eg_{\pi KK^*}(k_1+k_4)^\alpha
\frac{1}{(k_1-k_4)^2-m^2_{K^*}}\left[g_{\alpha\beta}-\frac{(k_1-k_4)_\alpha
(k_1-k_4)_\beta}{m^2_{K^*}}\right]\nonumber\\
&\times&[-(k_2+k_3)^\beta g^{\mu\nu}+(-k_1+k_2 +k_4)^\nu
g^{\beta\mu}+(k_1+k_3-k_4)^\mu
g^{\beta\nu}]\varepsilon_{3\mu}\varepsilon_{2\nu},\nonumber\\
{\cal M}^{sub}_{5b}&=&eg_{\pi KK^*}(2k_1+k_2)^\nu\frac{1}{(k_1+k_2)^2-m^2_K}
(k_1+k_2+k_4)^\mu\varepsilon_{3\mu}\varepsilon_{2\nu},\nonumber\\
{\cal M}^{sub}_{5c}&=&-eg_{\pi KK^*}g^{\mu\nu}\varepsilon_{3\mu}
\varepsilon_{2\nu},\nonumber\\
{\cal M}^{sub}_{6a}&=&-\sqrt{2}eg_{\rho KK^*}(-k_1-k_4)^\alpha
\frac{1}{(k_1-k_4)^2-m^2_\rho}\left[g_{\alpha\beta}-\frac{(k_1-k_4)_\alpha
(k_1-k_4)_\beta}{m^2_\rho}\right]\nonumber\\
&\times&[-(k_2+k_3)^\beta g^{\mu\nu}+(-k_1+k_2 +k_4)^\nu g^{\beta\mu}
+(k_1+k_3-k_4)^\mu g^{\beta\nu}]\varepsilon_{3\mu}\varepsilon_{2\nu},
\nonumber\\
{\cal M}^{sub}_{6b}&=&\sqrt{2}eg_{\rho KK^*}(2k_1+k_2)^\nu
\frac{1}{(k_1+k_2)^2-m^2_K}(k_1+k_2+k_4)^\mu\varepsilon_{3\mu}
\varepsilon_{2\nu},\nonumber\\
{\cal M}^{sub}_{6c}&=&-\sqrt{2}eg_{\rho KK^*}g^{\mu\nu}\varepsilon_{3\mu}
\varepsilon_{2\nu},\nonumber\\
{\cal M}^{sub}_{7a}&=&eg_{\rho KK^*}(-2k_1 +k_3)^\mu
\frac{1}{(k_1-k_3)^2-m^2_K}(k_1 -k_3 +k_4)^\nu\varepsilon_{3\mu}
\varepsilon_{2\nu},\nonumber\\
{\cal M}^{sub}_{7b}&=&-eg_{\rho KK^*}(2k_1+k_2)^\nu
\frac{1}{(k_1+k_2)^2-m^2_K}(k_1+k_2+k_4)^\mu\varepsilon_{3\mu}
\varepsilon_{2\nu},\nonumber\\
{\cal M}^{sub}_{7c}&=&2eg_{\rho KK^*}g^{\mu\nu}\varepsilon_{3\mu}
\varepsilon_{2\nu}.
\end{eqnarray}
As in our previous study of $\Theta^+$ production from photon-proton
reactions with four-body (or three-body if not considering decays of the
resonances in final state), we have introduced at the $KN\Theta$ vertex
another monopole form factor $F(({\bf p_1-p_3})^2)$ with cutoff 
parameter $\Lambda=0.5$ GeV and depending on the three momentum of $K^-$ 
besides the overall monopole form factor $F({\bf q}^2)$ with cutoff 
parameter $\Lambda=0.75$ GeV. As in the case of three-body final 
states, the cross sections for these reactions can also be expressed 
in terms of those for their subprocesses, similar to that given in 
Eq.(\ref{32cross}).

\subsection{results}\label{results}

\begin{figure}[ht]
\includegraphics[width=3.5in,height=4.5in,angle=-90]{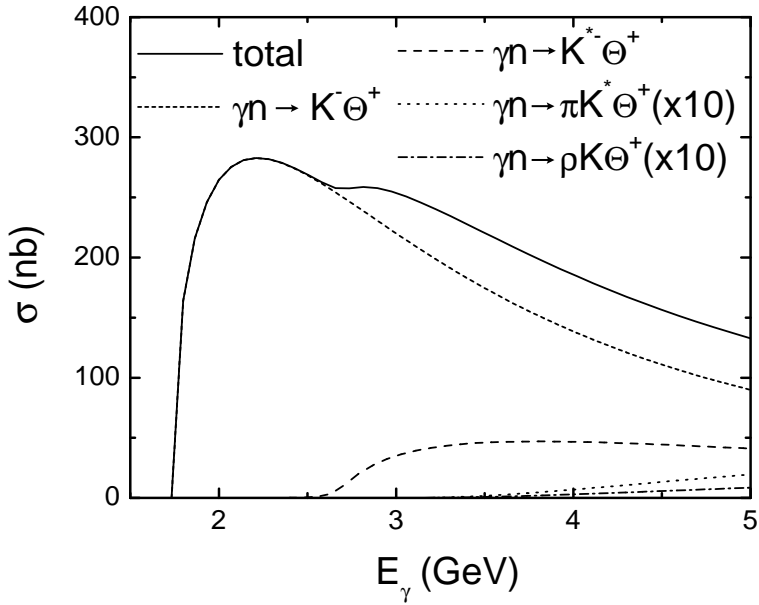}
\caption{Cross sections for $\Theta^+$ production from
photon-neutron reactions as functions of photon energy:
total (solid curve), $\gamma n\to K^-\Theta^+$ (short dashed curve),
$\gamma n\to K^{*-}\Theta^+$ (long dashed curve),
$\gamma n\to\pi^0K^{*-}(\pi^-\bar K^{*0})\Theta^+$ (dotted curve), and
$\gamma n\to\rho^0K^-(\rho^-\bar K^0)\Theta^+$ (dash-dotted curve).}
\label{cross1}
\end{figure}

In Fig.\ref{cross1}, we show the cross sections for $\Theta^+$ production
from photon-neutron reactions as functions as photon energy. The reaction 
$\gamma n\to K^-\Theta^+$ (short dashed curve) with two-body final state 
has the largest cross section with a peak value of about 280 nb at
$E_\gamma\sim 2.2$ GeV with main contribution from the diagram
involving $t$-channel $K^-$ exchange. We note that
this cross section is almost an order-of-magnitude larger than the
corresponding reaction in photon-proton reactions, i.e.,
$\gamma p\to \bar K^0\Theta^+$, due to the absence of a nucleon
propagator. The reactions $\gamma n\to\pi^0K^-\Theta^+$ and 
$\gamma n\to\pi^-\bar K^0\Theta^+$ with three-body final state, denoted 
combinedly as $\gamma n\to K^{*-}\Theta^+$ (long dashed curve)
in Fig.\ref{cross1}, have values of about 50 nb above $E_\gamma=3$ GeV.
The contributions from reactions with four-body final state are
smaller with only a few nb for both
$\gamma n\to\pi^0K^{*-}(\pi^-\bar K^{*0})\Theta^+$ (dotted curve),
and $\gamma n\to\rho^0K^-(\rho^-\bar K^0)\Theta^+$ (dash-dotted curve).
The total $\Theta^+$ production cross section from photon-neutron
reactions is given by the solid curve.

\section{discussions}
\label{discussions}

The calculated cross sections are sensitive to the value of cutoff
parameter used in form factors. If we increase the cutoff parameter
in the overall form factor by a factor of 2, i.e., from
$\Lambda=0.75$ GeV, which is based on our previous study of charmed
hadron production from photon-proton reactions \cite{liu2}, to 
$\Lambda=1.5$ GeV, then all the cross sections evaluated here for
photon energy near threshold will be increased by an order-of-magnitude.

\begin{figure}[ht]
\includegraphics[width=1.25in,height=1.5in,angle=0]{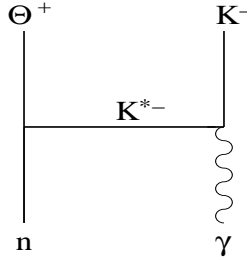}
\caption{Diagram for $\Theta^+$ production from photon-neutron
reactions via $K^*$ exchange.} \label{diagram4}
\end{figure}

Also, if we allow the $\Theta^+$ to also couple to $NK^*$, then it 
can be produced from the reaction $\gamma n\to K^-\Theta^+$
via $K^{*-}$ exchange as shown by the diagram in Fig.\ref{diagram4}.
In terms of the $K^*N\Theta$ interaction Lagrangian
\begin{eqnarray}
{\cal L}_{K^*N\Theta}&=&g_{K^*N\Theta}(\bar N\gamma_\mu\Theta
\bar K^{*\mu}+K^{*\mu}\bar\Theta\gamma_\mu N),
\end{eqnarray}
the amplitude for this reaction is given by
\begin{eqnarray}
{\cal M}_{\gamma n\to K^-\Theta^+}&=&g_{\gamma K^-K^{*-}}
g_{K^*N\Theta}\frac{F({\bf q}^2)}{t-m^2_{K^*}}
\epsilon_{\alpha\beta\mu\nu}p^\alpha_2
\epsilon_2^\beta(p_3-p_1)^\mu\bar\Theta(p_3)\gamma^\nu n(p_1),
\end{eqnarray}
Its cross section can be written similarly as in Eq.(\ref{cross}).

\begin{figure}[ht]
\includegraphics[width=3.5in,height=4.5in,angle=-90]{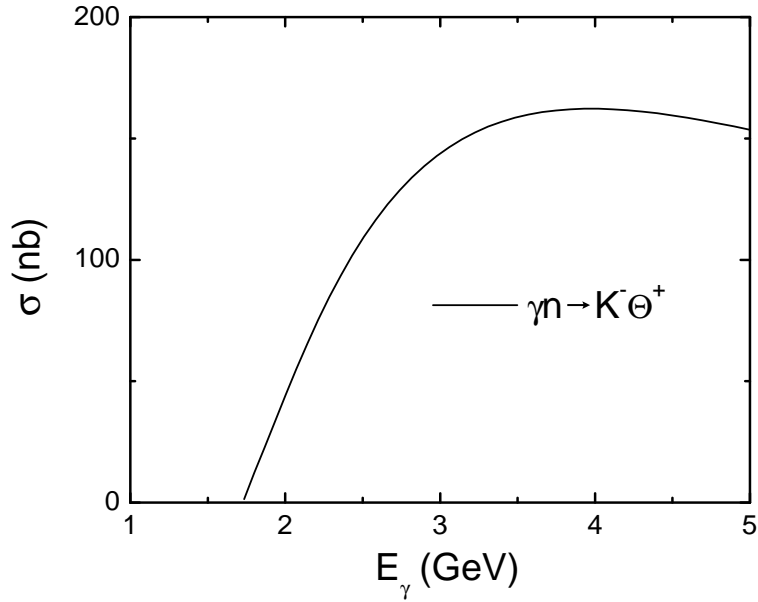}
\caption{Cross section for $\Theta^+$ production from photon-neutron
reactions via $K^*$ exchange as functions of photon energy.}
\label{cross2}
\end{figure}

The cross section for this reaction, using the same overall form factor with
cutoff parameter $\Lambda=0.75$ GeV as for other photon-neutron
reactions with two-body final states, depends on the unknown coupling constant
$g_{K^*N\Theta}$. If we take its value to be same as $g_{KN\Theta}$, i.e.,
$g_{K^*N\Theta}=4.4$, then the cross section for the reaction
$\gamma n\to K^-\Theta^+$ via $K^{*-}$ can reach 160 nb at
$E_\gamma\ge 3$ GeV as shown in Fig.\ref{cross2}. We note that although the
contribution from $K^{*-}$ exchange via the photon anomalous parity
interaction to $\Theta^+$ production cross section is smaller than
that due to $K^-$ exchange involving normal photon coupling in
photon-neutron reactions, it is important in photon-proton reactions
due to larger $g_{\gamma K^0K^{*0}}$ coupling than $g_{\gamma K^-K^{*-}}$
coupling. The cross section due to this process in photon-proton
reactions is more than 350 nb if $g_{K^*N\Theta}=4.4$ is used \cite{lkk},
and this is comparable to that measured in the SAPHIR experiment at 
Bonn University's ELSA accelerator.

\section{summary}
\label{summary}

The cross section for production of $\Theta^+$ baryon consisting
of $uudd\bar s$ quarks from photon-neutron reactions is evaluated in a
hadronic model that includes the $KN\Theta$ interaction with coupling
constant determined from the decay width of $\Theta^+$.
This model is based on a gauged SU(3) flavor symmetric Lagrangian with
the photon introduced as a $U_{\rm em}(1)$ gauged particle, which
has been used previously in studying production of strange baryons 
and charmed hadrons, after extending to $SU(4)$, in hadronic
reactions. Symmetry breaking effects are taken into account by 
using empirical hadron masses and coupling constants. Form factors 
of monopole type are introduced at interaction vertices to take into
account finite hadron sizes, and values of the cutoff parameters are
taken from fitting known cross sections of other reactions based on
similar hadronic models. It has already been used to evaluate the
cross sections for $\Theta^+$ production from meson-nucleon,
proton-proton, and photon-proton reactions \cite{liuko}.  The model is 
used in the present study to evaluate the cross section for $\Theta^+$
production from photon-neutron reactions. Including also the
anomalous parity coupling between photon and $K^*K$, it is found that the
dominant contribution is from the reaction $\gamma n\to K^-\Theta^+$
with two-body final state, and its cross section reaches a value of
280 nb. The one with three-body final state has 
a value of about 50 nb, and those involving four-body final states have 
cross sections of only a few nb. The large $\Theta^+$ production cross 
section in the reaction $\gamma n\to K^-\Theta^+$ than in the
corresponding reaction $\gamma p\to\bar K^0\Theta^+$ in photon-proton 
reactions, which has a peak value of only 40 nb \cite{liuko},
is due to the coupling of photon to $K^-$, which is absent in the 
photon-proton reactions.

We have also considered the process involving $K^{*-}$ exchange in the
reaction $\gamma n\to K^-\Theta^+$. Taking the unknown coupling constant
$g_{K^*N\Theta}$ to be the same as $g_{KN\Theta}$, this process 
increases the cross section by about 50\%. Because of the larger
$g_{\gamma K^0K^{*0}}$ coupling, which appears in the reaction
$\gamma p\to\bar K^0\Theta^+$, than $g_{\gamma K^-K^{*-}}$ coupling,
this process is more important in photon-proton reactions, 
and it may be responsible for the large cross section
measured in the SAPHIR experiment.

Since cross sections for $\Theta^+$ production in hadronic reactions
are proportional to the square of its coupling to $KN$ as in its
decay width to kaon and nucleon, it is then hard to understand the
large cross section measured in the SAPHIR experiment if the width
of $\Theta^+$ is less than 1 MeV as suggested in Ref.\cite{arndt}
based on reanalysis of $K^+n$ scattering data.

\begin{acknowledgments}
We thank Valeri Kubarovsky for communications. This
paper was based on work supported in part by the US National Science
Foundation under Grant No. PHY-0098805 and the Welch Foundation under
Grant No. A-1358.
\end{acknowledgments}

\end{document}